\documentclass{elsart}

\usepackage{natbib}

\usepackage{epsfig}

\usepackage{amssymb}

\begin{document}

\begin{frontmatter}

%%%%%%%%%%%%%%%%%%%%%%%%%%%%%%%%%%%%%%%%%%%%%%%%%%%%%%%%%%%%%%%%%%%%%%%%%%%
%%%%%%%%%%%%%%%%%%%%%%%%%%%%%%%%%%%%%%%%%%%%%%%%%%%%%%%%%%%%%%%%%%%%%%%%%%%
%%%%%%%%%%%%%%%%%%%%%%%%%%%%%%%%%%%%%%%%%%%%%%%%%%%%%%%%%%%%%%%%%%%%%%%%%%%
\title{
The SPICA coronagraphic instrument (SCI) for the study of exoplanets
}
%%%%%%%%%%%%%%%%%%%%%%%%%%%%%%%%%%%%%%%%%%%%%%%%%%%%%%%%%%%%%%%%%%%%%%%%%%%
%%%%%%%%%%%%%%%%%%%%%%%%%%%%%%%%%%%%%%%%%%%%%%%%%%%%%%%%%%%%%%%%%%%%%%%%%%%
%%%%%%%%%%%%%%%%%%%%%%%%%%%%%%%%%%%%%%%%%%%%%%%%%%%%%%%%%%%%%%%%%%%%%%%%%%%

\author[label1]{K. Enya\corauthref{cor},}
\author[label1]{T. Kotani,}
%%%-------------------------------------
\author[label2,label1]{K. Haze,}
\author[label3,label1]{K. Aono,}
%%%-------------------------------------
\author[label1]{T. Nakagawa,}
\author[label1]{H. Matsuhara,}
\author[label1]{H. Kataza,}
\author[label1]{T. Wada,}
\author[label1]{M. Kawada,}
\author[label1]{K. Fujiwara,}
\author[label1]{M. Mita,}
\author[label1]{S. Takeuchi,}
\author[label1]{K. Komatsu,}
\author[label1]{S. Sakai,}
%%%-------------------------------------
\author[label4]{H. Uchida,}
\author[label4]{S. Mitani,}
\author[label4]{T. Yamawaki,}
%%%-------------------------------------
\author[label5]{T. Miyata,}
\author[label5]{S. Sako,}
\author[label5]{T. Nakamura,}
\author[label5]{K. Asano,}
%%%-------------------------------------
\author[label6]{T. Yamashita,}
\author[label6]{N. Narita,}
\author[label6]{T. Matsuo,}
\author[label6]{M. Tamura,}
\author[label6]{J. Nishikawa,}
\author[label6]{E. Kokubo,}
%%%-------------------------------------
\author[label7]{Y. Hayano,}
\author[label7]{S. Oya,}
%%%-------------------------------------
\author[label8]{M. Fukagawa,}
\author[label8]{H. Shibai,}
%%%-------------------------------------
\author[label9]{N. Baba,}
\author[label9]{N. Murakami,}
%%%-------------------------------------
\author[label10]{Y. Itoh,}
\author[label11]{M. Honda,}
\author[label12]{B. Okamoto,}
\author[label13]{S. Ida,}
\author[label14]{M. Takami,}
\author[label15]{L. Abe,}
\author[label16]{O. Guyon,}
\author[label17]{P. Bierden,}
\author[label18]{T. Yamamuro}\\

%%%%%%%%%%%%%%%%%%%%%%%%%%%%%%%%%%%%%%%%%%%%%%%

\address[label1]{Institute of Space and Astronautical Science, 
      Japan Aerospace Exploration Agency, \\
      3-1-1 Yoshinodai, Chuo-ku, Sagamihara, Kanagawa 252-5210, Japan}
\address[label2]{The Graduate University for Advanced Studies, 
       3-1-1 Yoshinodai, Sagamihara, Tyuou-ku, Kanagawa 252-5210, Japan}
\address[label3]{Department of Physics, Graduate School of Science, 
       University of Tokyo, Bunkyo-ku, Tokyo 113-0033, Japan}
\address[label4]{Aerospace Research and Development Directorate,
       Japan Aerospace Exploration Agency, Tsukuba Spcae Center, 
       2-1-1 Sengen, Tsukuba-shi, Ibaraki 305-8505, Japan}
\address[label5]{Institute of Astronomy, School of Science, 
       University of Tokyo, 2-21-1 Osawa, Mitaka, Tokyo 181-0015,Japan}
\address[label6]{National Astronomical Observatory of Japan, 2-21-1 Osawa, 
       Mitaka, Tokyo 181-8588, Japan}
\address[label7]{Subaru Telescope, National Astronomical Observatory of Japan, 
       650 North A'ohoku Place, Hilo, Hawaii96720, U.S.A.}
\address[label8]{Department of Earth and Space Science, Graduate School of Science, 
       Osaka University, 1-1Machikaneyama, Toyonaka, Osaka 560-0043, Japan}
\address[label9]{Division of Applied Physics, Faculty of Engineering, 
       Hokkaido University, Sapporo 060-8628, Japan}
\address[label10]{Graduate School of Science and Technology, Kobe University, 
       1-1 Rokkodai, Nada, Kobe 657-8501, Japan}
\address[label11]{Department of Information Sciences, Kanagawa University, 
       2946 Tsuchiya, Hiratsuka, Kanagawa, 259-1293, Japan}
\address[label12]{Institute of Astrophysics and Planetary Sciences, 
       Faculty of Science, Ibaraki University, 2-1-1 Bunkyo,Mito, 
       Ibaraki 310-8512, Japan}
\address[label13]{Tokyo Institute of Technology, Ookayama, Meguro-ku, 
       Tokyo 152-8551, Japan}
\address[label14]{Institute of Astronomy and Astrophysics, Academia Sinica. 
       P.O. Box 23-141, Taipei 10617, Taiwan, R.O.C}
\address[label15]{Laboratoire Fizeau, Universit\'e  de Nice Sophia-Antipolis, 
       Parc Valrose, 06108 Nice Cedex 02, France}
\address[label16]{Subaru Telescope, National Astronomical Observatory of Japan, 
       650 North A'ohoku Place, Hilo, Hawaii96720, U.S.A.}
\address[label17]{Boston Micromachines Corporation,
       30 Spinelli Place, Cambridge, MA 02138, U.S.A}
\address[label18]{Optcraft Corporation, 3-16-8 Higashi-Hashimoto, 
       Sagamihara, Midori-ku, Kanagawa 252-0144, Japan}

%%%%%%%%%%%%%%%%%%%%%%%%%%%%%%%%%%%%%%%%%%%%%%%%%%%%
\corauth[cor]{Corresponding author.\\
{\it Email address:} enya@ir.isas.jaxa.jp (K. Enya)}
%%%%%%%%%%%%%%%%%%%%%%%%%%%%%%%%%%%%%%%%%%%%%%%%%%%%

%%%%%%%%%%%%%%%%%%%%%%%%%%%%%%%%%%%%%%%%%%%%%%%%%%%%%%%%%%%%%%%%%%%%%%%%%%%
%%%%%%%%%%%%%%%%%%%%%%%%%%%%%%%%%%%%%%%%%%%%%%%%%%%%%%%%%%%%%%%%%%%%%%%%%%%
%%%%%%%%%%%%%%%%%%%%%%%%%%%%%%%%%%%%%%%%%%%%%%%%%%%%%%%%%%%%%%%%%%%%%%%%%%%
%%%%%%%%%%%%%%%%%%%%%%%%%%%%%%%%%%%%%%%%%%%%%%%%%%%%%%%%%%%%%%%%%%%%%%%%%%%
\begin{abstract}

We present the SPICA Coronagraphic Instrument (SCI), which has been 
designed for a concentrated study of extra-solar planets (exoplanets).
SPICA mission provides us with a unique opportunity to make 
high contrast observations
because of its large telescope aperture, the simple pupil shape, 
and the capability for making infrared observations from space.
The primary objectives for the SCI are the direct coronagraphic  
detection and spectroscopy of Jovian exoplanets in infrared, 
while the monitoring of transiting planets is another important target.
The specification and an overview of the design of the instrument are shown.
In the SCI, coronagraphic and non-coronagraphic modes are applicable
for both an imaging and a spectroscopy.
The core wavelength range and the goal contrast of the coronagraphic 
mode are 3.5--27$\mu$m, and 10$^{-6}$, respectively.
Two complemental designs of binary shaped pupil mask coronagraph
are presented.
The SCI has capability of simultaneous observations 
of one target using two channels, a short channel with an InSb detector
and a long wavelength channel with a Si:As detector.
We also give a report on the current progress in the development of key 
technologies for the SCI.

\end{abstract}
%%%%%%%%%%%%%%%%%%%%%%%%%%%%%%%%%%%%%%%%%%%%%%%%%%%%%%%%%%%%%%%%%%%%%%%%%%%
%%%%%%%%%%%%%%%%%%%%%%%%%%%%%%%%%%%%%%%%%%%%%%%%%%%%%%%%%%%%%%%%%%%%%%%%%%%
%%%%%%%%%%%%%%%%%%%%%%%%%%%%%%%%%%%%%%%%%%%%%%%%%%%%%%%%%%%%%%%%%%%%%%%%%%%
%%%%%%%%%%%%%%%%%%%%%%%%%%%%%%%%%%%%%%%%%%%%%%%%%%%%%%%%%%%%%%%%%%%%%%%%%%%

%%%%%%%%%%%%%%%%%%%%%%%%%%%%%%%%%%%%%%%%%%%%%%%%%%%
%%%%%%%%%%%%%%%%%%%%%%%%%%%%%%%%%%%%%%%%%%%%%%%%%%%
%%%%%%%%%%%%%%%%%%%%%%%%%%%%%%%%%%%%%%%%%%%%%%%%%%%
%%%%%%%%%%%%%%%%%%%%%%%%%%%%%%%%%%%%%%%%%%%%%%%%%%%

\end{frontmatter}

\parindent=0.5 cm

\clearpage

%%%%%%%%%%%%%%%%%%%%%%%%%%%%%%%%%%%%%%%%%%%%%%
\section{Background and Scientific Objective}
%%%%%%%%%%%%%%%%%%%%%%%%%%%%%%%%%%%%%%%%%%%%%%

We regard the systematic study of extra-solar planets (exoplanets) to be one 
of the most important tasks to be undertaken in space science in the near future. 
The enormous contrast between the parent star and the planet presents 
us with a serious problem.
Therefore, there is a requirement for special instruments using 
techniques specifically designed to improve the contrast 
in order to perform a systematic observation of exoplanets.
There are a number of different techniques currently used to observe exolanets.
Since the first report by \citet{Mayor1995} using the radial velocity method, 
more than 450 exoplanets have been discovered. 
It has also been shown that observations monitoring the transits of exoplanets provide a 
valuable means for studying them \citep{Charbonneau2000}. 
Not only detection but also spectroscopic studies of some transiting exoplanets 
have been 
carried out (e.g., \citet{Deming2005};  \citet{Tinetti2007}; \citet{Swain2010}). 
Recently, the spatially resolved direct detection of an exoplanet by coronagraphic 
imaging was finally reported (e.g., \citet{Marois2008}; \citet{Kalas2008}). 
While these methods are quite valuable, the current targets of these methods 
tend to be strongly biased. 
Many of the targets for radial velocity observations and the monitoring of  
transiting planets are ”Hot-Jupiters” that are close to their parent stars. 
On the other hand, exoplanets observed by current coronagraphic imaging 
tend to be strongly limited to young, giant outer planets. 
So the coronagraphic observation of mature, outer planets is still 
missing from the population. 
Considering this situation, 
we believe that the next important step in this 
field is a systematic study of exoplanets of various ages, masses, 
and distances from their parent stars. 
A spectroscopic survey of known exoplanets is especially important for 
characterizing planetary atmospheres.

The Space Infrared telescope for Cosmology and Astrophysics (SPICA) is an 
astronomical mission optimized for mid-infrared (MIR) and 
far-infrared astronomy with a 3m class 
($\sim$3.0m effective pupil diameter/$\sim$3.2m physical diameter
in the current design) 
on-axis telescope cooled to $<$\,6K \citep{Nakagawa2010}. 
We have proposed to develop a Coronagraphic Instrument for SPICA (SCI) 
and to carry out essential study in exoplanet science with the SCI
(P.I.: K. Enya; \citet{Enya2010} and its references).

SPICA has advantages as the platform for the SCI: 
it is perfectly free  from band pass limitations and wavefront turbulence 
caused by the atmosphere.
The cryogenic telescope provides high sensitivity in the infrared region.
High stability is expected as the cryogenic telescope is to be launched into 
deep space, the Sun-Earth L2 Halo orbit.
In addition, the structure of the SPICA telescope, adopting a monolithic 
primary mirror and carefully designed secondary support, yields a
clean point spread function (PSF).

The major scientific tasks are described below. 
We regard that the instrument should 
be designed to carry out two ``critical tasks''; 
one is the coronagraphic detection 
and characterization of exoplanets, 
and, the other is the monitoring of transiting planets. 
There are other important scientific tasks which can be carried out  
with the ``given''  performance of the designed instrument.

\begin{figure}[!h]
\begin{center}
\includegraphics*[width=16cm]{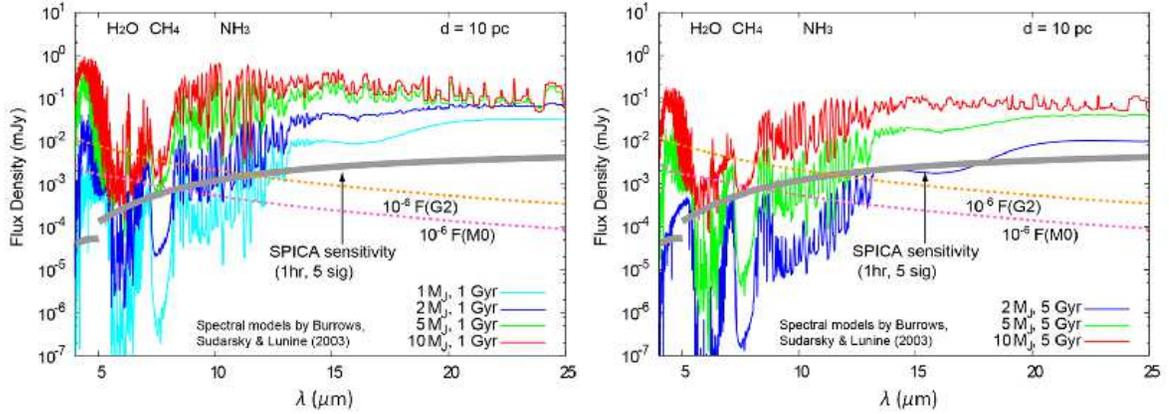}
\end{center}
\caption{
Calculated SEDs of 1\,Gyrs (left) 
and 5\,Gyrs (right) old Jovian planets 
with various masses presented as Fig.\,13 in \citet{Burrows2003}, 
and properties relating to SPICA observations. 
10pc is assumed as the distance to the planetary system. 
The gray solid curve shows the sensitivity limit of imaging with SPICA. 
The orange and purple dashed lines show the scaled SEDs of G2 and M0 type stars, 
respectively.
These figures are from \citet{Fukagawa2009}.
}
\label{figure1}
\end{figure}

\begin{enumerate}

\item
\underline{Coronagraphic observations}:\\
One of the two critical scientific tasks for the SCI is
to pursue the direct detection and spectroscopy of exoplanets.
The SCI suppresses light from the parent star, and in spectroscopy mode 
working with coronagraph reveals 
essentially important spectral features in the MIR region, 
CH$_4$, H$_2$O, CO$_2$, CO, NH$_3$. 
Jovian exoplanets around (1) nearby stars($\leq$\,10pc) and (2) young and moderately 
old stars ($\leq$\,1Gyr old) are the primary targets for coronagraphic observations
with the SCI (Fig.\ref{figure1}). 
The former are suitable for detailed spatially-resolved observations, 
and the latter for gaining an understanding of the history of the formation 
of the planetary system. 
With the SCI we expect to be able to create an atlas of the various spectra of 
exoplanets by making observations on $\sim$100s of targets.
More detail is given in \citet{Fukagawa2009}, \citet{Matsuo2011}. 
\\

\item
\underline{Monitoring transiting planets}:\\ 
The other critical scientific task is precise monitoring 
of transiting planets in order to characterize their 
spectral features.
Coronagraphic observations by the SCI cover the ``outer planets'', 
i.e., those at $\sim$10AU or further from the parent star. 
Therefore, the observation of transiting planets and 
coronagraphic observations are complementary techniques. 
It should be noted that the method for monitoring transiting planets  
requires the planets to be in edge-on orbits, 
and therefore discovery before observation with SPICA.   
The SCI has many advantages as a transit monitoring instrument.
It has the capability for simultaneous observations with two detectors, 
having filters optimized for the spectral features of planetary atmospheres.
Of the focal plane instruments on the SPICA mission, only the SCI covers
the spectrum down to wavelengths of $\sim$1$\mu$m 
(coronagraphic high contrast images are guaranteed only for wavelengths
down to 3.5$\mu$m; however the instrument has an InSb detector which is sensitive 
at shorter wavelengths).  
Another advantages of the SCI for this task are
the superior pointing stability in 
the SPICA instruments, 0.05arcsec can be 
realized by an internal sensor, 
and potential capability for defocusing 
to avoid saturation by using an internal 
deformable mirror.
Additionally, partial readout of the detectors can be
adopted to improve the exposure/readout duty cycle when imaging.
\\

\item
\underline{Other observations of planetary systems}:\\
Some other important observations relating to extra-solar planetary systems 
are planned with the given instrument design.
Color Differential Astrometry (CDA) is being considered for the observation 
of planetary systems \citep{Abe2009}.
CDA is a challenging method, but it has the advantage of having the capability 
of observing ``inner planets'' and does not require planets in edge-on orbits.
Observations of the ``snow line'' (e.g., \citet{Honda2009}) 
and other features in circumstellar disks are other important targets 
for the SCI (\citet{Tamura2009}; \citet{Takami2010}). 
\\

\end{enumerate}

By using the spectral data sets of various exoplanets obtained by the SCI, 
we expect that our understanding of the whole planetary system will be 
significantly improved.

%%%%%%%%%%%%%%%%%%%%%%%%%%%%%%%%%%%%
\section{Instrument}
%%%%%%%%%%%%%%%%%%%%%%%%%%%%%%%%%%%%

%==================================
\subsection{Specification}
%==================================

The specification of the instrument is summarized in Table\,\ref{table1}. 
An overview of the current optical design of the SCI is shown in Fig.\ref{figure2}. 
The requirement that gives us the limiting short wavelength (3.5$\mu$m) 
is derived for the direct detection and spectroscopy 
of Jovian exoplanets. 
As shown in Fig.\,\ref{figure1}, it is expected that the Spectral Energy 
Distribution (SED) of Jovian exoplanets has a peak in the 3.5--5$\mu$m 
wavelength region (Burrows2003). 
So this wavelength region is one of the most appropriate regions for the 
direct detection of Jovian planets. 
Wavelength coverage from MIR to 3.5$\mu$m allows us to study interesting
molecular features of the planetary atmospheres, 
e.g., H$_2$O, CH$_4$, NH$_3$.
The goal contrast (10$^{-6}$ at PSF) is derived for the systematic
study of Jovian planets, including not only extremely massive, young
planets but lower mass planets (down to $\sim$1M$_J$ with 1Gyr old targets)
and older planets (up to 5Gyr old massive targets).
It should be noted that the PSF subtraction technique has the potential 
to improve the final contrast after image processing, 
e.g., by one order or more (\cite{Trauger2007}; \cite{Haze2009}).
Although the diffraction limited image at a wavelength of 5$\mu$m is a specification 
of the SPCIA telescope, wavefront correction using a deformable 
mirror (DM), make observations at shorter 
wavelengths possible with the SCI. 
The SCI is primarily used for coronagraphic imaging and in coronagraphic 
spectroscopy mode. 
On the other hand, non-coronagraphic imaging and spectroscopy are also possible 
because the coronagraph mask can be removed. 
In non-coronagraphic mode the SCI is useful as a general purpose fine-pixel camera 
and a spectrometer. 

The high contrast in the coronagraphic image is guaranteed only within a field 
determined by the inner working angle (IWA), the outer working angle (OWA),
and the number of actuators in the DM, whereas the SCI has a FoV of 1'$\times$1'.
The number of DM actuators is 32$\times$32 which
provides the capability to control the wavefront over 
an area of $16\lambda/D \times 16\lambda/D$ in the coronagraphic image.
This high contrast area, $16\lambda/D \times 16\lambda/D$, determines
the limit to the outer working angle (OWA) in the design of the coronagraph
shown in Table.\,\ref{table1}.

Two channels, a short and a long wavelength 
channel, have been adopted, together with a beam splitter. 
The short and long wavelength channels have an InSb 
and a Si:As detector, respectively, and cover
the wavelength regions $\leq$5$\mu$m and  $\geq$5$\mu$m.
This two channel design has some advantages:
First, a higher sensitivity for the 3.5--27$\mu$m wavelength region is 
obtained compared to a single channel design.
Secondary, the simultaneous observation of one target with two 
channels is possible.
Thirdly, an appropriate pixel scale can be determined for each channel.
In the current design, 1.5$\times$ Nyquist sampling 
(i.e., over sampling) at 3.5$\mu$m and 5$\mu$m
has been adopted for the short and long channels, respectively.
The simultaneous observation of one target with two 
channels is possible.

The IWA of the SCI, 3.3$\lambda/D$, corresponds to $\sim$1 and  
$\sim$2\,arcsec at wavelengths of 5$\mu$m and  10$\mu$m, respectively.
These values are no smaller than those of the next generation
of ground based coronagraphs, e.g.,the Gemini Planet Imager (GPI)
or coronagraphs for 30m class telescopes.  
However, such coronagraphs used with ground based telescopes mainly
work only in the window of atmospheric transmittance in infrared. 
This suggests that ground based coronagraphs are useful
for imaging young planets, and are complementary to the 
SCI's capability for wide band spectroscopy for cooler exoplanets.
In comparison with the JWST coronagraphs,
the most important advantage of the SCI is its capability for 
spectroscopic measurements which JWST does not have.
Furthermore, the contrast of the SCI is significantly higher 
owing to the active wavefront control.
The capability of the simultaneous use of two channels in the SCI
is useful for image subtraction of coronagraphic observations, 
and monitoring transiting planets with wide 
wavelength coverage, which is not possible with JWST.
\cite{Matsuo2011} presents more on the comparison between the SCI
and the JWST coronagraphs, especially on the point of detectability
of outer cooled planets.

%==================================
\subsection{Overview of the Instrument }
%==================================

\begin{figure}[!h]
\begin{center}
\includegraphics*[width=16cm]{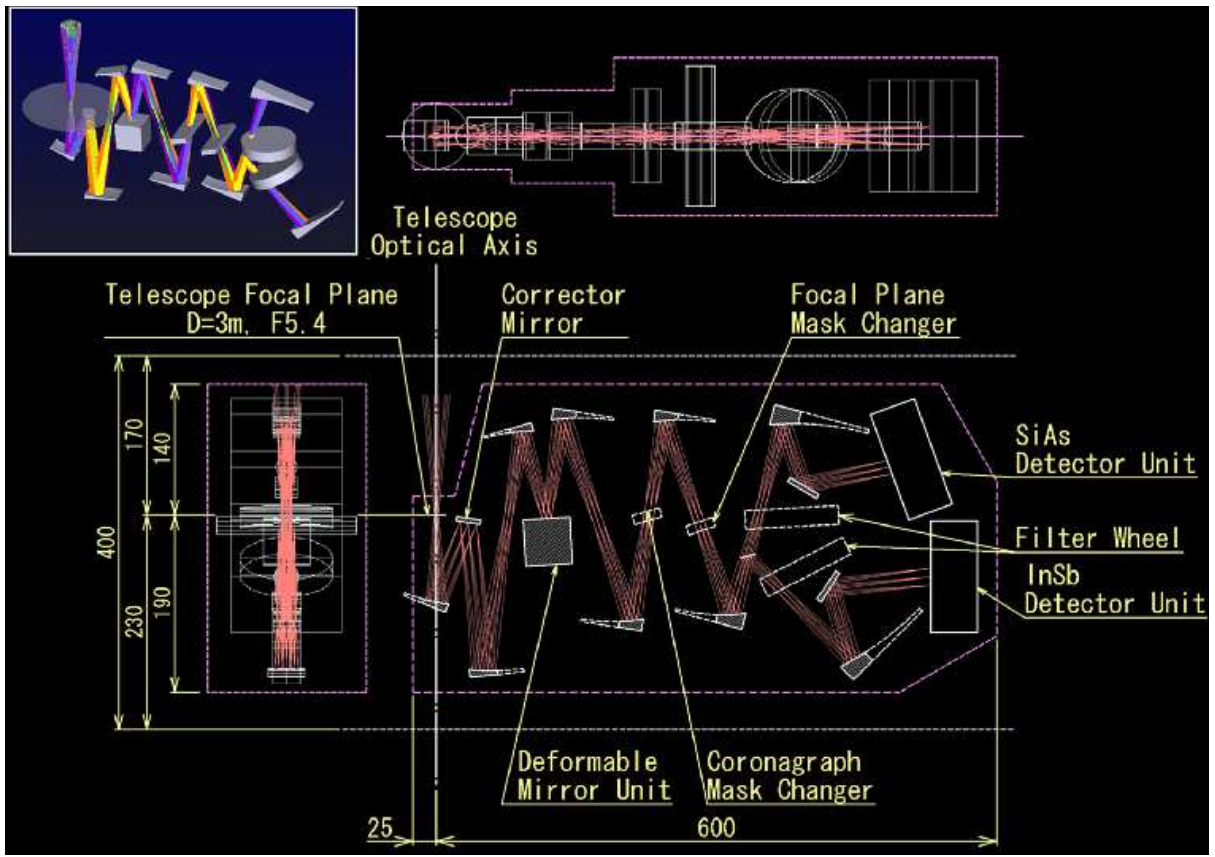}
\end{center}
\caption{Overview of the optical design of the SCI.
}
\label{figure2}
\end{figure}

\begin{figure}[!h]
\begin{center}
\includegraphics*[width=9cm]{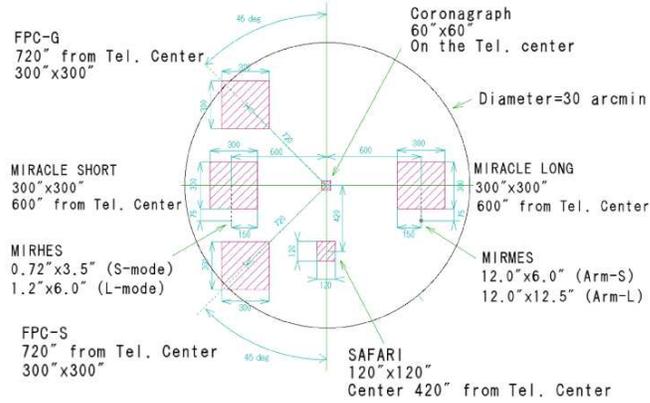}
\end{center}
\caption{Current design of the distribution of the FoV for the SPICA instruments. 
}
\label{figure3}
\end{figure}

The optics of the SCI is compact 
with the optical axis in one plane (Fig.\,\ref{figure2}). 
The total mass is $\sim$30kg
and the whole instrument, together with the telescope, is cooled to 5K
before operation to enable the SPICA to achieve ultra-high sensitivity 
in the far- and mid-infrared regions. 
The SCI uses the center of the field of view in the focal plane of the 
telescope to obtain the best wavefront accuracy and 
symmetricity of the obscuration by the secondary mirror and its 
support structure(Fig.\,\ref{figure3}). 
All the mirrors for collimation and focusing in the SCI are off-axis parabolas. 
Aberration is minimized at the center of the 
FoV in order to obtain the best coronagraphic PSF of the parent star. 
The pre-coronagraphic optics and the coronagraphic optics are common 
for both wavelength channels, 
while the post-coronagraphic optics is split into two channels.

The pre-coronagraphic optics consists of reflective optical devices in order to 
avoid ghosts and/or a wavelength dependent refractive indices which are sometimes 
issues in lens optics. 
In contrast to the other focal plane instruments on the SPICA mission,
the SCI has no pick-off mirror, with the beam from the telescope secondary mirror
arriving directly on the collimating mirror (Fig.\,\ref{figure2}).
This configuration makes the SCI compact under the constraint presupposing
no lens optics.
Minimizing the number of mirrors reduces the total light scattering 
at the surfaces of the mirrors. 
The first mirror of the SCI is placed after the focal point of the telescope. 
This is convenient for optically testing the SCI and the telescope. 
A DM is included in the SCI to correct the wavefront errors of the telescope.

As shown in Fig.\ref{figure4}, 
the baseline solution for the coronagraphic method is a binary-shaped 
pupil mask (see Sec.\ref{sec_crg}).
With this method, only one pupil mask modifies the PSF and provides
the required contrast. 
A focal plane mask is used to obscure the 
bright core in the coronagraphic PSF and to prevent scattered 
light from the PSF core polluting the dark region of the coronagraphic
PSF in the post-coronagraphic optics. 
Interchangeable focal plane masks will be used to realize
different observing modes, including a slit that can be used both 
with and without the pupil mask to provide spectroscopic capability
in coronagraphic and non-coronagraphic observing modes.
Fig.\,\ref{figure5} shows the concept of spectroscopy 
working together with coronagraphy in the SCI.

The post-coronagraphic optics uses a beam splitter to split the optical path 
into two channels, the short wavelength channel with an InSb detector 
and the long wavelength channel with a Si:As detector.
Each channel has filter wheels, and simultaneous observation of the same target 
with the two channels is possible. 
Each filter wheel contains a transmissive disperser (e.g., a grism) 
for spectroscopy.

In order to reduce technical risk in the development, 
another design of the instrument without the use of a DM is also 
under consideration, together with the full-equipped solution 
described above.
As the result of this simplification, the contrast at the PSF 
is limited to be $\sim$10$^{-4}$.
In this case, the advantage of high contrast over JWST is basically lost.
However, the spectroscopy working with coronagraph 
remain a unique capability of the SCI.  
It is already know that there are young outer planets observed
by direct imaging (e.g., \citet{Marois2008}).
These planets are enough bright in infrared for the spectroscopic
observation using the simplified design of the SCI.
We believe that the spectrum data of such targets for the wide 
infrared wavelength range is essentially unique and important.

%%%%%%%%%%%%%%%%%%%%%%%%%%%%%%%%%%%%%%%%%%%%
\section{Key technologies}
%%%%%%%%%%%%%%%%%%%%%%%%%%%%%%%%%%%%%%%%%%%%

The SCI requires challenging technologies to realize high contrast 
for both imaging and spectroscopy over a wide MIR wavelength range.
One of the critical technologies is the design, development and manufacture 
a coronagraph which can yield a high contrast PSF.
We focused on a coronagraph using a multiple 1-dimensional barcode 
mask \citep{EnyaAbe2010}
which is a type of binary shaped pupil 
mask (e.g., \cite{Kasdin2005}; \cite{Vanderbei2004}; \cite{Tanaka2006}).
Another key technology is the adaptive optics, which improves the
wavefront and stability of the PSF because the specification for these
is more critical for the SCI
than for the SPICA.
Optical ghosts and scattering in the SCI should be enough nullified 
practically to avoid polluting the high contrast image.
With this in mind, the development of a cryogenic MIR chamber 
for end-to-end tests of the coronagraph are ongoing.
More details are presented in the following sections.

%===========================================
\subsection{Coronagraph}\label{sec_crg}
%===========================================

A coronagraph to produce high contrast images is the core function of the SCI. 
The coronagraph for SPICA has to work over a wide MIR wavelength region
in a cryogenic environment, so a coronagraph without transmissive optical devices 
is preferable. 
The coronagraph should be insensitive to telescope pointing errors caused by 
vibration from the cryogenic cooling system and the attitude control system. 
Achromatism (except the scaling effect for the size of the PSF) 
is also an important property. 
The coronagraph should be applicable for the pupil of the SPICA, which is partly obscured 
by the secondary mirror and its support structure. 
After taking these points into consideration, a coronagraph using 
a binary shaped pupil mask was selected as the primary method for our study. 
Experiments were carried out
to confirm the feasibility of this strategy.

\begin{figure}[!h]
\begin{center}
\includegraphics*[width=9.5cm]{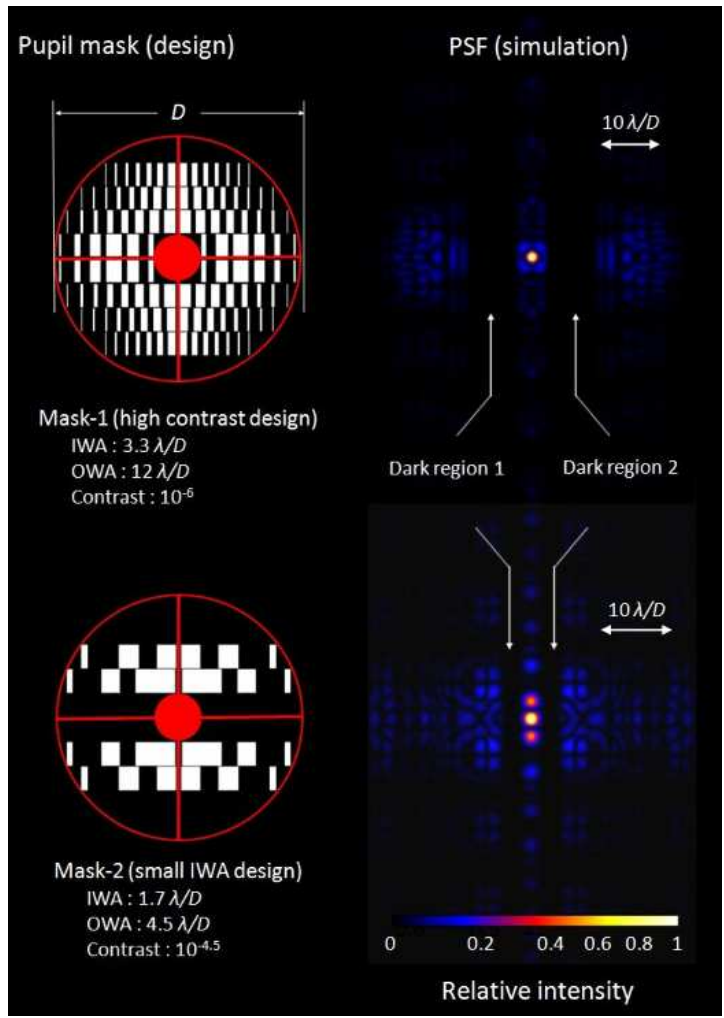}
\end{center}
\caption{
Design of a multi-barcode 1D coronagraph 
for the SPICA pupil.
Top and bottom panels show current baseline and
an optional design, respectively.
The mask design (left) and the simulated PSF using 
this pupil (right). 
Transmissivity of the mask is 1 and 0 at white and
black area, respectively.
}
\label{figure4}
\end{figure}

\begin{figure}[!h]
\begin{center}
\includegraphics*[width=14cm]{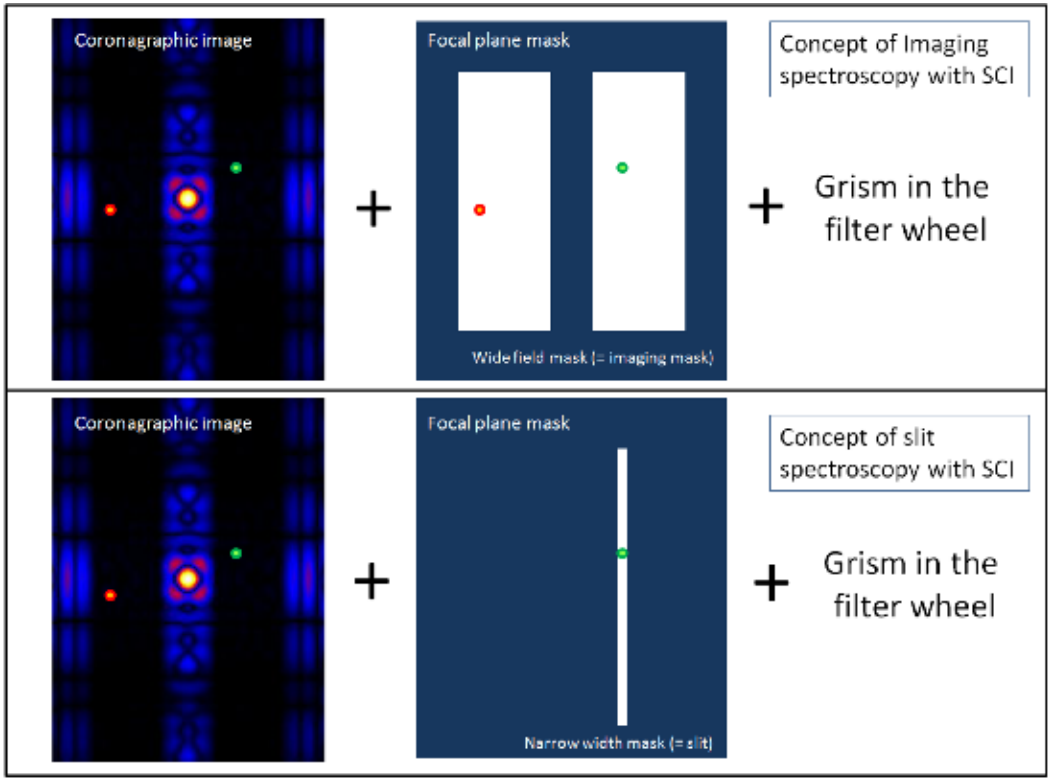}
\end{center}
\caption{Spectroscopy working together with 
coronagraphy in the SCI. 
}
\label{figure5}
\end{figure}

\begin{figure}[!h]
\begin{center}
\includegraphics*[width=16cm]{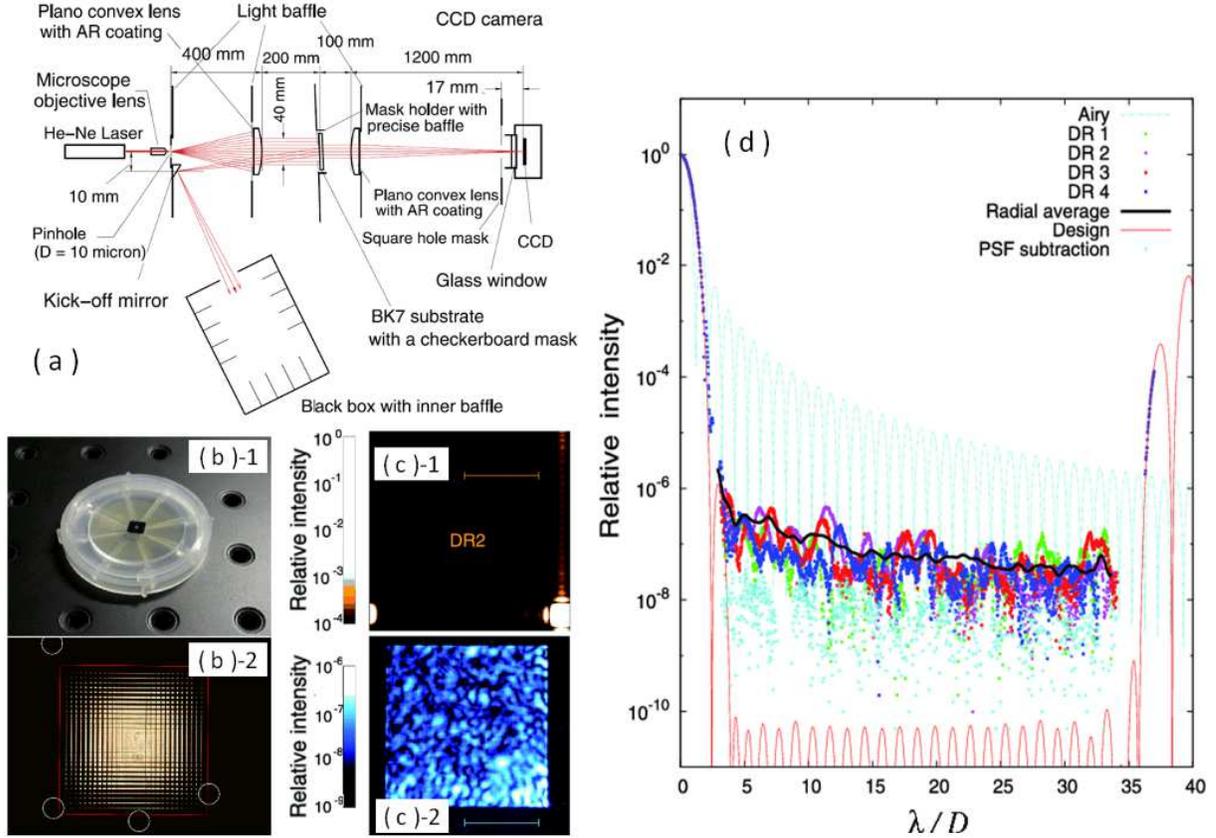}
\end{center}
\caption{
(a): Configuration used in demonstrating the operation 
of a binary shaped pupil mask.
(b): Checkerboard mask manufactured on a glass substrate by 
nano-fabrication technology using electron beam lithography 
and a lift-off process. Top and bottom panels show 
whole of the device, and microscopic image, respectively.
This mask is designed to produce four dark region of square shape,
DR1-DR4, around the core of the PSF.  
(c): Experimental PSF (top) and high sensitivity image 
of a dark region, DR2,  of the PSF (bottom). 
(d): The observed and theoretical coronagraphic 
profiles as well as the theoretical Airy profile. 
Each profile is normalized to the peak intensity in each image.
}
\label{figure6}
\end{figure}

\begin{figure}[!h]
\begin{center}
\includegraphics*[width=16cm]{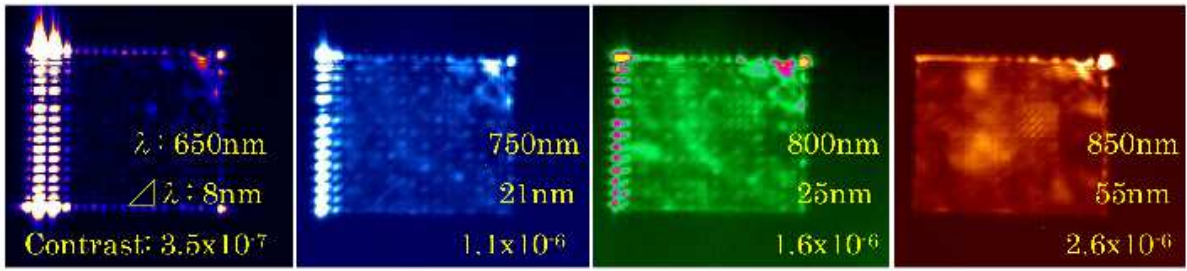}
\end{center}
\caption{
Results of a multi-color/broadband experiment using a checkerboard mask. 
Only images of the dark region are shown. 
Ghosting caused by the lens puts a practical limit on the contrast 
at longer wavelengths. 
}
\label{figure7}
\end{figure}

\begin{figure}[!h]
\begin{center}
\includegraphics*[width=16cm]{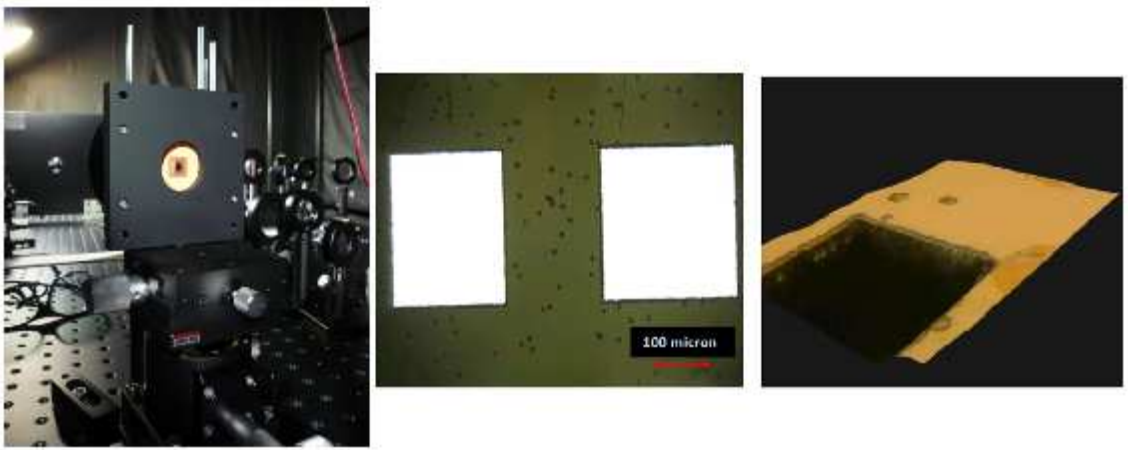}
\end{center}
\caption{
Left: a free standing checkerboard mask made of a thin copper plate
installed in the test setup using a visible laser in air.
Middle and right: microscope pictures of the free standing checkerboard 
mask made of thin copper plate.
}
\label{figure8}
\end{figure}

\begin{itemize}

\item
\underline{Demonstration of the principle}: \\
First we carried out experiments to validate 
the performance of the coronagraph with a checkerboard mask, which is a
type of binary shaped pupil mask, using a visible laser in an air ambient. 
Checkerboard masks consisting of 100nm thick aluminum patterns on BK7 glass substrates 
were constructed using nano-fabrication technology. Electron beam lithography 
and a lift-off process were used in the fabrication process. 
Optimization of the checkerboard masks was executed using the LOCO software presented 
by \citet{Vanderbei1999}. 
The contrast, derived by averaging over the dark region using a linear scale
and comparing this with the peak intensity, 
was 6.7$\times$10$^{-8}$ (Fig.\,\ref{figure6}-(c),(d)).
\\

\item
\underline{Multi-band experiment with non-zero bandwidth light sources}: \\
We installed multi-band Super Luminescent Diode Light 
Sources (SLED) for this experiment, and confirmed that the binary shaped 
pupil mask coronagraph works for light sources with center 
wavelengths ($\lambda$)of 650, 750, 800, 850nm with bandwidths ($\Delta\lambda$)
of 8, 21, 25, and 55nm, respectively. 
The results of this experiment are shown in Fig.\,\ref{figure7}.
More details are given in \citet{Haze2010}. 
\\

\item
\underline{Mask design for the SPICA pupil}: \\
We designed some binary shaped pupil masks for the SPICA 
pupil mask having the obscuration shown in Fig.\,\ref{figure4}.
These designs consist of multi-barcode masks and have coronagraphic power 
in one dimension only. 
As a result, large opening angle is realized
with keeping to satisfy the specification for the IWA.
In the masks shown in Fig.\,\ref{figure4},
mask-1 is the baseline design.
Additionally, mask-2 provides a small IWA to explore the field 
closer to the parent star.
On the other hand, the contrast of mask-2 is not 
as high as mask-1 because there is a trade-off between the IWA and the contrast.
These two masks are complementary, and can be changed
by a mechanical mask changer.
It should be noted that the principle of the barcode mask was 
presented by \citet{Kasdin2005}.
More details are presented in \citet{EnyaAbe2010}
\\

\item
\underline{Free standing masks:} \\
The trial fabrication of free standing masks (i.e., masks without a substrate) 
was carried out using various manufacturing methods. 
We tested the coronagraphic performance using a manufactured free standing 
mask made of thin copper plate and a visible He-Ne laser (Fig.\,\ref{figure8}). 
This mask was also used in an experiment to demonstrate wavefront correction, 
and this was found to work successfully with this mask (see Sec.\,\ref{sec_wfc}).
\\

\end{itemize}

Combining these results implies that it is reasonable to assume
that our binary shaped pupil mask coronagraph will work at MIR wavelengths. 
The development of a cryogenic chamber for end-to-end demonstration of the 
MIR coronagraph is ongoing (see Sec.\,\ref{sec_milct}). 
Toughness tests with vibration and acoustic load 
is to be undertaken for the mask in future work.
\\

%===========================================
\subsection{Adaptive optics}
%===========================================

The specification for the wavefront quality of the SPICA telescope is 350\,nm rms, 
while the requirement for the SCI is higher by a factor of 10 or more. 
Therefore, the SCI needs an internal active wavefront correction system which works 
at the temperature of the SCI (i.e., 5K). 
Therefore, we began the development of a cryogenic DM, 
which is one of the critical components needed to realize the SCI.

\begin{figure}[!h]
\begin{center}
\includegraphics*[width=16cm]{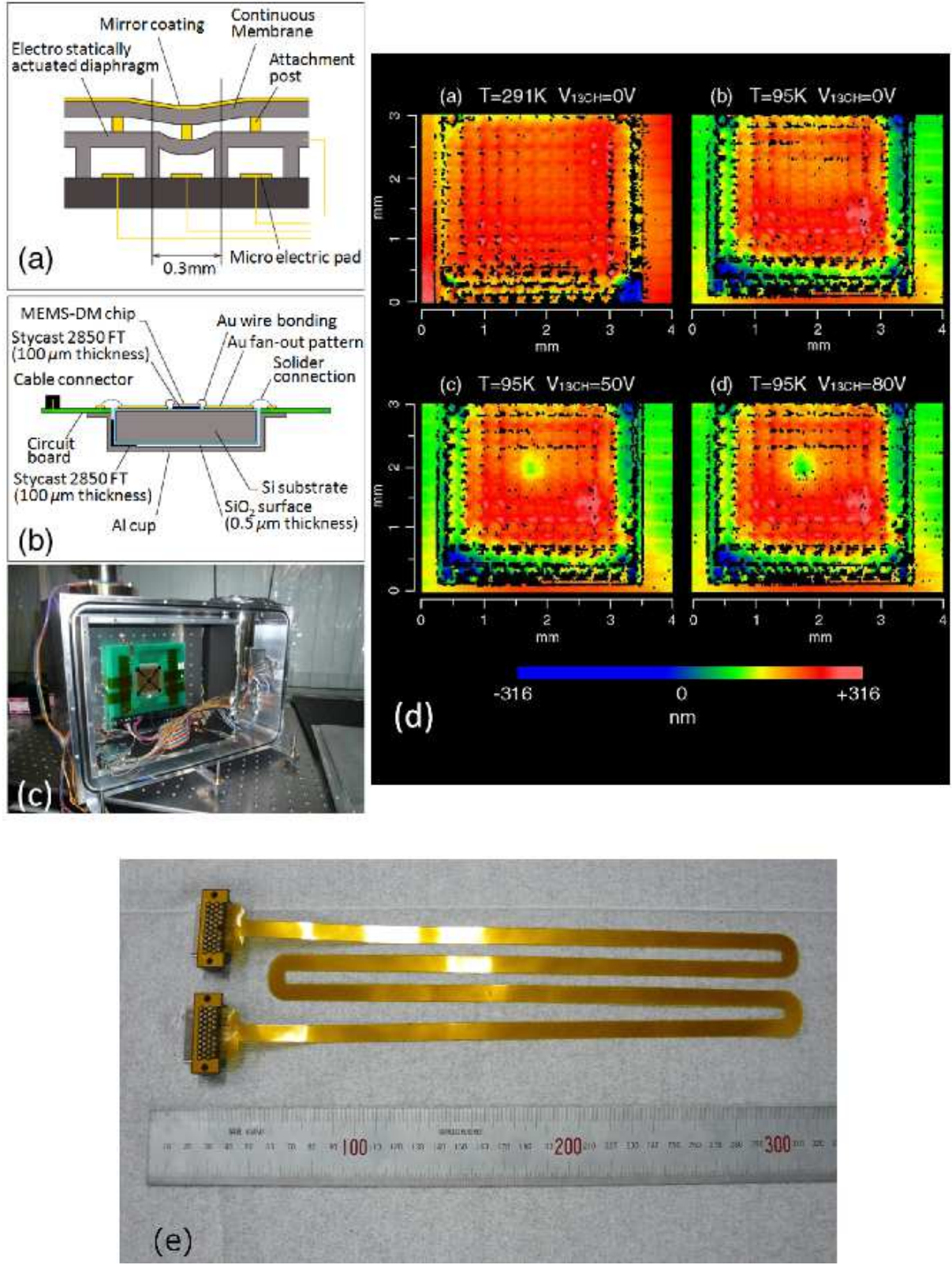}
\end{center}
\caption{
(a)-(c): schematic views and a photograph of the prototype cryogenic DM unit. 
(d): results of a demonstration of the prototype cryogenic DM. 
(e): A sample of the film print cable.  
}
\label{figure9}
\end{figure}

\begin{itemize}

\item
\underline{Demonstration with a prototype device}: \\
We developed a prototype DM with 32 channels 
consisting of a Micro Electro Mechanical Systems (MEMS) chip fabricated 
by the Boston Micromachines Corporation (BMC) and a special 
silicon substrate designed to minimize the thermal stress caused by cooling. 
Fig.\,\ref{figure9} shows the result of a demonstration with this prototype 
in a cryogenic environment. 
For the first step of this work, a liquid 
nitrogen cooled chamber was used (i.e., without liquid helium) for convenience.
As a result, the lowest temperature of the DM was limited to 95K.
More than 80\% of the linear thermal shrinkage of the silicon 
when cooling from room temperature to 5K occurs 
in the temperature range from 293 to 95K \citep{Okaji1999}.
This fact suggests that the development strategy starting is reasonable.
For the full demonstration of the cryogenic DM,
development of a cryogenic 32$\times$32 channel DM and a 5K chamber 
is ongoing.
More details are given in \citet{Enya2009}.
\\

\item
\underline{Structural analysis of a 1000 channel DM}:\\ 
From the studies with the prototype, we found that the thermal stress is 
the critical issue for our cryogenic DM if we intend to go to a larger format, 
e.g., a 1000 channel DM. 
Therefore, we carried out simulation studies of the thermal deformation and stress 
induced by cooling in order to obtain design solutions that would maintain 
the flatness of a 1000 channel DM cooled to 5K. 
We obtained an initial design, and further improvement is ongoing.
 \\

\item
\underline{Development of a film print cable}: \\
Parasitic heat passing through the cables between the cooled DM and the warm 
electronics is another important issue. 
In the SPICA mission, the allocated parasitic heat for all 
focal plane instruments is limited to only 10mW.
 We attempted the manufacture of film print cables in order to reduce 
the parasitic heat to an ultra-low level ( $<$1mW via 1000 cables). 
We confirmed that the first version of our film print cable is sufficiently 
thin and provides electrical insulation up to 200V (Fig.\,\ref{figure9}).  \\

\end{itemize}

These results suggest that the development of a cryogenic DM with 1000 
channels is viable and that it will work at 5K. 
We are also considering a cryogenic multiplexer as a backup solution 
to the film print cables. 
Testing of toughness of the cryogenic DM will be also undertaken in future work.

%=========================================
\subsection{Wavefront correction}\label{sec_wfc}
%=========================================

\begin{figure}[!h]
\begin{center}
\includegraphics*[width=16cm]{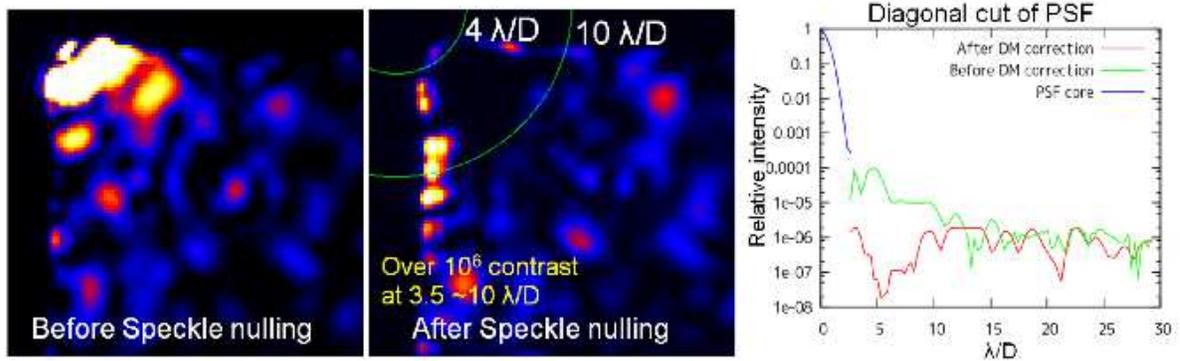}
\end{center}
\caption{
Results of the speckle nulling experiment. 
Left and middle: images of a dark region of the PSF 
obtained with a square hole mask,
before and after speckle nulling, respectively. 
Right: Diagonal cut through the PSF.
}
\label{figure10}
\end{figure}

Another important issue is how to operate a DM in order 
to correct wavefront errors. 
To develop an algorithm for SPICA/SCI, we initiated experiments using 
a visible laser, a commercially-available DM from BMC,
a free standing coronagraph mask having the checkerboard design, 
and a CCD camera.
Coronagraphic images were taken, and the speckle nulling method
was applied to cancel the speckle in a dark region of the 
PSF \citep{Malbet1995}.  
A remarkable improvement of the coronagraphic contrast, 
to become much better than 10$^{-6}$ , was confirmed for an area 
just outside the inner working angle, which is an important 
area in the search for exoplanets (Fig.\,\ref{figure10}).
The algorithm used in this experiment included hundreds of 
iterations to produce anti-speckles.
We plan in future work to try a more sophisticated algorithm
which requires less number of images with different phases 
as presented in \cite{Borde2006}, \cite{Giveon2007}.  
For the SCI, the phase shift will be achieved by operating an internal
device (i.e., a deformable mirror) or moving the secondary
mirror of the SPICA telescope.
More detail is given in \cite {Kotani2010}.

%=========================================
\subsection{Cryogenic MIR Testbed}\label{sec_milct}
%=========================================

Our previous laboratory experiments on the coronagraph were performed 
at room temperature and atmospheric pressure, and at visible wavelengths, 
whereas the SPICA coronagraph will have to be evaluated at cryogenic 
temperatures, in a vacuum, and at infrared wavelengths. 
In order to complete end-to-end testing of the MIR coronagraph at 5K,
we are developing a cryogenic vacuum chamber
in the Institute of Space and Astronautical Science (ISAS) of
the Japan Aerospace Exploration Agency (JAXA).
The cryogenic DM and free-standing mask 
described above will be included in this chamber.

%=========================================
\subsection{Internal optics structure}
%=========================================

It is planned that the internal off-axis mirror, its support
structure, and the optical bench in the SCI will be made 
of the same material in order to avoid deformation of the
optics due to mismatching of the thermal
expansion coefficients.
The primary candidate for this material is aluminum.
Determination of the specification for the mirror quality
is ongoing.
It is expected that our requirements for the quality of 
the mirror surface are more relaxed than those for
optical coronagraph missions targeting
terrestrial planets \citep{Shaklan2006}
because of the differences in target contrast and
the observation wavelength region.

As presented above, key technologies required to realize a coronagraph
with spectroscopic capabilities in the MIR for SPICA are being assessed.
The indications are that all the major issues, including the binary shaped pupil 
mask coronagraph for the MIR and the cryogenic adaptive optics, can be resolved.

%%%%%%%%%%%%%%%%%%%%%%%%%%%%%%%%%%%%
\section*{Acknowledgments}
%%%%%%%%%%%%%%%%%%%%%%%%%%%%%%%%%%%%

We deeply thank and pay our respects to all the pioneers in this field, 
especially R. Vanderbei and J. Kasdin. 
This work is supported by JAXA.

%%%%%%%%%%%%%%%%%%%%%%%%%%%%%%%%%%%%

%%%%%%%%%%%%%%%%%%%%%%%%%%%%

\clearpage

%%%%%%%%%%%%%%%%%%%%%%%%%%%%
%Table
%%%%%%%%%%%%%%%%%%%%%%%%%%%%

\begin{center}

\begin{table}
\begin{center}
\caption{Specification for the SCI}
\begin{tabular}{ll}
\hline
Wavelength ($\lambda$)* & Core wavelength lambda = 3.5-27$\mu$m \\
                      & \hspace{10mm} Short wavelength channel: $\lambda$ = 3.5-5$\mu$m \\
                      & \hspace{10mm} Long wavelength channel: $\lambda \ge 5 \mu$m \\
Coronagraph method    & Binary shaped pupil mask \\ 
Observation mode      & Coronagraphic imaging \\
                      & Coronagraphic spectroscopy \\
                      & Non-coronagraphic imaging \\
                      & Non-coronagraphic spectroscopy \\ 
contrast              & $10^{-6}$ @PSF \\ 
Inner working angle (IWA)   & 3.3$\lambda/D$ **\\ 
Outer working angle (OWA)   & 16$\lambda/D$ \\ 
Filter                & Band-pass filters for each channel \\ 
Detector              & a 1K $\times$ 1K Si:As array for the long wavelength channel\\ 
                      & a 1K $\times$ 1K InSb array for the short wavelength channel \\ 
FoV                   & High-contrast coronagraphic FoV:16$\lambda/D$ \\
                      & (FoV of $1' \times 1'$ is available but high-contrast is not guaranteed\\
                      & out of 16$\lambda/D$)\\ 
Spectral resolution   & $\sim$20 and $\sim$200 (realized by transmissive dispersion devices,\\
                      & e.g. grisms) \\ 
\hline
\end{tabular}
\label{table1}
\end{center} 
* At $\lambda < 3.5\mu$m, high contrast imaging is not guaranteed,
but the instrument has sensitivity through the InSb detector.\\
** $D$ is telescope aperture diameter.
\end{table}
\end{center}

\clearpage

%%%%%%%%%%%%%%%%%%%%%%%%%%%%
%Figures
%%%%%%%%%%%%%%%%%%%%%%%%%%%%

%\vspace*{5mm}

%%%%%%%%%%%%%%%%%%%%%%%%%%%%%%%%%%%%%%%%%%%%%
%%%%%%%%%%%%%%%%%%%%%%%%%%%%%%%%%%%%%%%%%%%%%

\vspace{10mm}

%%%%%%%%%%%%%%%%%%%%%%%%%%%%%%%%%%%%%%%%%%%%%
%%%%%%%%%%%%%%%%%%%%%%%%%%%%%%%%%%%%%%%%%%%%%

\clearpage

%%%%%%%%%%%%%%%%%%%%%%%%%%%%%%%%%%%%%%%%%%%%%
%%%%%%%%%%%%%%%%%%%%%%%%%%%%%%%%%%%%%%%%%%%%%

%%%%%%%%%%%%%%%%%%%%%%%%%%%%%%%%%%%%%%%%%%%%%
%%%%%%%%%%%%%%%%%%%%%%%%%%%%%%%%%%%%%%%%%%%%%

%%%%%%%%%%%%%%%%%%%%%%%%%%%%%%%%%%%%%%%%%%%%%
%%%%%%%%%%%%%%%%%%%%%%%%%%%%%%%%%%%%%%%%%%%%%

%%%%%%%%%%%%%%%%%%%%%%%%%%%%%%%%%%%%%%%%%%%%%
%%%%%%%%%%%%%%%%%%%%%%%%%%%%%%%%%%%%%%%%%%%%%

%%%%%%%%%%%%%%%%%%%%%%%%%%%%%%%%%%%%%%%%%%%%%
%%%%%%%%%%%%%%%%%%%%%%%%%%%%%%%%%%%%%%%%%%%%%

%%%%%%%%%%%%%%%%%%%%%%%%%%%%%%%%%%%%%%%%%%%%%
%%%%%%%%%%%%%%%%%%%%%%%%%%%%%%%%%%%%%%%%%%%%%

%%%%%%%%%%%%%%%%%%%%%%%%%%%%%%%%%%%%%%%%%%%%%
%%%%%%%%%%%%%%%%%%%%%%%%%%%%%%%%%%%%%%%%%%%%%

%%%%%%%%%%%%%%%%%%%%%%%%%%%%%%%%%%%%%%%%%%%%%
%%%%%%%%%%%%%%%%%%%%%%%%%%%%%%%%%%%%%%%%%%%%%

%%%%%%%%%%%%%%%%%%%%%%%%%%%%%%%%%%%%%%%%%%%%%
%%%%%\begin{figure}[!h]
%%%%% \begin{center}
%%%%% \includegraphics*[width=8cm]{fig_milct.eps}
%%%%% \end{center}
%%%%% \caption{
%%%%% Top: cryogenic chamber for testing the MIR coronagraph 
%%%%% under construction in ISAS/JAXA.
%%%%% Bottom: temperature curves obtained from a cooling tests of the chamber. 
%%%%% }
%%%%% \label{figure10}
%%%%% \end{figure}
%%%%%%%%%%%%%%%%%%%%%%%%%%%%%%%%%%%%%%%%%%%%%

%%%%%%%%%%%%%%%%%
 \end{document}